\documentclass[10pt]{amsart}
\usepackage{amssymb}
\usepackage{enumerate}
\usepackage[utf8]{inputenc}
\usepackage{graphicx}
\usepackage{hyperref}
\usepackage[export]{adjustbox}
\begin{document}
\begin{abstract}
We explore the effects coming from the nested structures on propagation of radial light rays going through them in the simple self-similar configuration containing pairs of voids and overdense regions. Each pair can be described with the Lemaitre-Tolman-Bondi metric and is composed of interior dust ball and exterior Schwarzschild region. A relation between the redshift and amount of nested pairs is studied numerically in order to answer a question whether a strong relativistic effect can be obtained when every component of the structure can be described by the weak field approximation of general relativity.
\end{abstract}
\title[Optical properties of self-similar structure]{Optical properties of self-similar structure in Lemaitre-Tolman-Bondi model}
\author{Jarosław Kopiński}
\address{
Institute of Theoretical Physics, Faculty of Physics, University of Warsaw, Pasteura 5, 02-093 Warsaw, Poland}
\maketitle
\section*{Introduction}
The question whether small-scale inhomogeneities in our Universe have significant effect on its large scale (the so-called 'backreaction problem') still remains open. On the one hand, Green and Wald argue that the Friedman-Lemaître-Robertson-Walker (FLRW) model is sufficient to describe the real structures and the discrepancies which come from the assumption about isotropy are small and can be neglected \cite{WalIsh}. Another viewpoint is such that the latter model is insufficient for the proper description of the reality and the phenomena of accelerated expansion is an artefact coming from the unjustified matching of the experimental data to an oversimplified model \cite{sysraks}. Third group consists of researchers that admit that small-scale inhomogeneities has some non-negligible effect, but it cannot be used to completely rule out the famous cosmological constant term in the Einstein field equations \cite{Buchert}. The backreaction problem has its origins in the simple fact that the procedure of taking an average is not well defined in general relativity. There have been some attempts to introduce a covariant way to describe inhomogeneous distribution of matter and energy in some region by the effective quantities, but none of them provides a fully approved way of doing that.
\\
\indent
There are many different possibilities to study the inhomogeneous structures. One of them is to make use of so-called Swiss-cheese models. They exploit the possibility of matching two different solutions of the Einstein field equations. The construction of such models starts from taking a homogeneous space and replacing some regions with a vacuum solutions, with matching condition imposed on some sphere. A slight variation of this procedure will be used to construct the multiscale structure in this paper.
\\
\indent
We follow the conventions of \cite{krasinski, gravitation}; in particular, we use a system of units for which $c = G = 1$.
\section{Lemaitre-Tolman-Bondi metric}
\indent
This chapter is dedicated to review some properties of the Lemaitre-Tolman-Bondi (LTB) metric, which will be used to construct self-similar structure. It is a natural generalization of Schwarzschild and FLRW metrics, and contains them as a special cases \cite[Chapter 18]{krasinski}. The explicit form given in standard coordinates is
\begin{equation} \label{eq: metr}
g = -\mathrm{d}t^2 + \frac{R_{,r}^2}{1 + 2 E} \mathrm{d}r^2 + R^2 \, \mathrm{d}\Omega^2,
\end{equation}
with functions $R(r,t)$ and $E(r)$. The $\mathrm{d} \Omega$ is a metric of the unit sphere expressed in standard angles $\{\theta, \phi\}$.
\\
\indent
Einstein field equations with metric of the form (\ref{eq: metr}) and vanishing cosmological constant are equivalent to the following system
\begin{equation} \label{eq: efe}
  \begin{aligned}
  G_{tt}& = \big( RR^2_{,t} -2ER \big) _{,r}=4 \pi \rho R^2R_{,r}, \\
  G_{rr}& = 2 \frac{R_{,tt}}{R} + \bigg( \frac{R_{,t}}{R} \bigg)^2 - \frac{2E}{R^2} =0, \\
  G_{rt}& = 2 R_{,[rt]} =0, \\
   G_{\phi \phi} &=G_{\theta \theta} = \frac{R_{,tt}}{R} + \frac{R_{,t}R_{,rt}}{RR_{,r}}+\frac{R_{,rtt}}{R_{,r}}-\frac{E_{,r}}{RR_{,r}}=0. \\
  \end{aligned}
\end{equation}
In order to manipulate (\ref{eq: efe}) we have to justify the possibility of dividing equations by $R_{,r}$ and $R_{,t}$. As it turns out, the vanishing of $R,_t$ everywhere leads to the Nariai solution with the cosmological constant \cite{nariai} or to the constant density case, which is not interesting from our point of view. On the other hand, vanishing of the radial derivative implies the existence of the singular points (physical or coordinate system singularities) and leads to the Datt-Ruban solution \cite[Chapter 19]{krasinski}, which will not be treated here. 
\\
\indent
After rewriting second equation from (\ref{eq: efe}) to the form
\begin{equation}
\left(R(R_{,t})^2-2ER \right)_{,t}=0,
\end{equation}
and introducing an integration constant $M=M(r)$ we have
\begin{equation} \label{eq: baseq}
 (R_{,t})^2=2E+\frac{2M}{R}.
\end{equation}
Function $M(r)$ has a clear physical interpretation as a mass contained in a ball of radius $r$. This is due to the integral expression, 
\begin{equation} \label{eq: defm}
M(r_0)=\int_{0}^{r_0} 2 \pi \rho R^2R_{,r}dr + M(0)=\int_{0}^{R(r_0)} 2 \pi \rho x^2 dx + M(0),
\end{equation}
where, using the regularity property in $r=0$ \cite[Chapter 18]{krasinski}, $M(0)=0$. It is worth to note that the expression (\ref{eq: defm}) is similar to the flat space definition of mass as an integral of density over a ball but differs from it by the volume element.
\\
\indent
The last two equations from (\ref{eq: efe}) are trivially satisfied (assuming that the derivatives $\partial_r$ and $\partial_t$ commutes), so the system reduces to (\ref{eq: baseq}) and (\ref{eq: defm}) -- a definition of mass. Function $R$ is so-called area radius, named after one of the coordinate from the Schwarzschild metric, whereas $E$ is the local generalization of curvature parameter $k$ in the FLRW models. 
\section{Self-similar structure}
\begin{figure}
\centering
\includegraphics[scale=0.4]{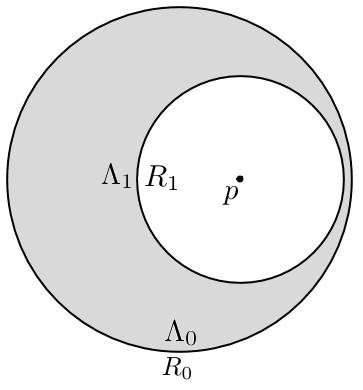}
\caption{Second stage of building the self-similar structure.}  \label{fig: fig1}
\end{figure}

We begin the construction by choosing the sign of local curvature $E$ to be positive. The first step is the same as in \cite{korz}. We consider spherically symmetric model which consists of internal ball of dust and external Schwarzschild region, matched on a spherical surface $r=R_0$. Using the connection between those two metrics and an LTB one, we claim that this situation can be described by the following choice of function $E$ and $M$ in the equation (\ref{eq: baseq}),
\begin{equation} \label{eq: defem}
E(r)=\begin{cases}
E_{0}r^2& \quad  r<R_0 \\
-\frac{M_0}{r}& \quad r\geq R_0
\end{cases},\qquad 
M(r)=\begin{cases}
\frac{4}{3}\pi \rho r^3 \quad \  &r<R_0 \\
M_0  \quad  &r\geq R_0
\end{cases},
\end{equation}
where $E_0$ is the curvature coefficient of the internal ball of dust, $\rho$ is the density of matter inside it and $M_0$ is the Schwarzschild mass parameter in the external void. Definitions (\ref{eq: defem}) can be simplified after the natural matching condition on the surface $r=R_0$. In that case we have
\begin{equation} \label{eq: defem2}
E(r)=\begin{cases}
-\frac{M_0}{R_0^3}r^2 \quad &r<R_0 \\
-\frac{M_0}{r} \quad  &r\geq R_0
\end{cases},\quad
M(r)=\begin{cases}
M_0 \left( \frac{r}{R_0} \right)^3 \quad  &r<R_0 \\
M_0 \quad  &r\geq R_0
\end{cases}.
\end{equation}
Under those assumptions, the initial slice of spacetime ($t=0$) is describing the interior FLRW part and exterior spherically symmetric static region, which is the part of Schwarzschild spacetime (Birkhoff's theorem, \cite[Chapter 32]{gravitation}). One can ask whether the effects coming from the nonlinear character of the Einstein equations will be important in the description of this system. We can argue that if the strength of gravitational field, measured by the compactness parameter $\epsilon$,
\begin{equation}
\epsilon = \frac{M_0}{R_0},
\end{equation}
will be small, then the approximation using linearized gravity can be used. Another way of saying that is to consider the expansion of exterior metric in the compactness parameter,
\begin{equation}
g= -\mathrm{d}t^2+ \mathrm{d}x^{i} \mathrm{d}x_{i} + \epsilon h_1 + O(\epsilon^2),
\end{equation}
where $h_1$ is the first correction term. We say that a weak field limit is the case when $\epsilon \ll 1$ and the strong field limit when $\epsilon \approx 1$. An observer placed outside the ball of dust can determine whether the $\epsilon$ is big or small by measuring the redshift of light ray coming from the center.
\\
\indent
In order to proceed with the construction of a self-similar structure, we want to describe the dust ball from the perspective of the matching conditions of two metrics, FLRW and Schwarzschild one. In the interior region, we introduce the hyperspherical coordinates on $S^3$, with the metric of the form
\begin{equation}
g_{I} = a^2 \big[ \mathrm{d}\lambda ^2 + \sin^2 \lambda (\mathrm{d}\theta^2 + \sin^2 \theta \, \mathrm{d}\phi^2)\big]. 
\end{equation}
In those coordinates, the ball of dust can be described with the $\lambda \in [0, \Lambda_0]$.
Surface $\lambda=\Lambda_0$ can be matched to the appropriate sphere $r=R_0$ in the Schwarzschild metric using the Israel matching condition \cite{israel}. Firstly, from the equality of metrics coefficient on the boundary sphere we have
\begin{equation} \label{eq: match1}
R_0 = a \sin \Lambda_0.
\end{equation}
Another condition is coming from the equality of the second fundamental forms on this surface, namely
\begin{equation} \label{eq: match2}
\sqrt{2\frac{M_0}{R_0}} = \sqrt{2 \epsilon} = \sin \Lambda_0.
\end{equation}
Equations (\ref{eq: match1}) and (\ref{eq: match2}) imply the connection between the Schwarzschild mass and the boundary angle $\Lambda_0$,
\begin{equation}
M_0 = \frac{a \sin^3 \Lambda_0}{2}.
\end{equation}
\\
\indent
Now we proceed to the second step of construction. We select a $\Lambda_1 < \Lambda$, point $p$ in the internal dust region and consider a new coordinate system with the center in point $p$, which can be chosen in an arbitrary way because of the isotropy of interior solution. We want to excise a ball of dust with the origin in the point $p$ and the angular diameter $2\Lambda_1$. After that, another matching can be performed, this time with the dust in the exterior region and Schwarzschild part in the interior. It is possible to use the Israel matching conditions again because they are invariant with the change of the position of the two matching solutions. The conditions on the surfaces $\lambda=\Lambda_1$, $r=R_1$ (Fig. \ref{fig: fig1}) read
\begin{equation}
\begin{aligned}
M_1&=\frac{a\sin^3 \Lambda_1 }{2}=M_0\bigg(\frac{ \sin\Lambda_1}{\sin\Lambda_0}\bigg)^3, \\
R_1&= R_0\bigg(\frac{\sin\Lambda_1}{\sin \Lambda_0}\bigg).
\end{aligned}
\end{equation}
The next step is to select radius, $R_2 < R_1$ and replace the interior Schwarzschild solution with another dust region, with center $p$ and radius $R_2$. In principle, value of $R_2$ can be chosen arbitrarily, but in order for the whole structure to be self-similar we require that
\begin{equation}
R_2=R_0 \left( \frac{\sin \Lambda_1}{\sin \Lambda_0} \right)^3.
\end{equation}
The self-similarity of the structure means that the compactness parameter $\epsilon$ is the same for the both dust regions as seen from the surrounding vacuum. This concludes the description of elementary blocks from which the whole structure will be made. The next step would be to choose another point, $p_1$, inside the smaller dust region and excise a ball with the center in this point  and diameter $\Lambda_1$ as measured in the hyperspherical coordinate system with origin at $p_1$. In general, we can repeat those step go generate $N$ pairs of Schwarzschild and FLRW regions embedded inside each other in a self-similar way defined above. Corresponding masses and radii are given by
\begin{equation}
\begin{aligned}
M_{n} = M_0 \left( \frac{\sin \Lambda_1}{\sin \Lambda_0} \right)^{3n}, \\
R_{n+1} = R_0 \left( \frac{\sin \Lambda_1}{\sin \Lambda_0} \right)^{3n},
\end{aligned}
\end{equation} 
for $n \geq 1$, such that $\frac{M_n}{R_{n+1}}=\epsilon$.
\subsection*{Evolution of the structure}
The construction above is given on the initial hypersurface $t=0$ of the LTB metric. Each component is connected with adjacent ones with the Israel matching conditions on the boundary between them. In order for this construction to be valid after the initial moment of time, those conditions has to be conserved for $t>0$. We can show that by moving to description of every pair of Schwarzschild and FLRW regions with LTB metrics, $\{g_n\}_{n \in N}$ separately. If those metrics are continuous with the first derivatives, then the Israel matching conditions are valid everywhere. To show that this is true, we first consider equation (\ref{eq: baseq}) for every $g_n$. After the differentiation with respect to $t$, it reads
\begin{equation}
R,_{tt} = -\frac{M}{R^2}.
\end{equation} 
Together with the initial conditions $R(0,r)$ and $R,_{t}(0,r)$ it determines the full continuous evolution of $R$. Another function which depends on $t$ in $g_n$ is the $R,_r$. To show that this is also continuous for every $t$ we consider the differentiation of (\ref{eq: baseq}) with respect to $r$. We get a ordinary partial differential equation for $R,_r$,
\begin{equation}
R,_{rt} = - \frac{M R,_r} {R^2 R,_{t}},
\end{equation}
which, along with initial conditions, is also well defined and describes the appropriate evolution of $R_r$.
\\
\indent
This is only a part of Israel matching conditions. Another issue is the compatibility of second fundamental forms. Radial vector which is perpendicular to every matching surface has the form $\nu = \frac{\sqrt{1+2E}}{R,_r}\partial_r$. The goal is to check whether Lie derivatives of metrics $q_n$ (induced on every boundary surface from $g_n$'s) with respect to $\nu$ are continuous functions for every $t$. This statement is true because
\begin{equation}
\mathcal{L}_{\nu}q_n = 2R \sqrt{1+2E} \mathrm{d} \Omega, 
\end{equation}
which contains only continuous functions.
\\
\indent
Using the arguments above, we can see that the evolution of structure is well defined and will proceed as follows -- every dust region will evolve accordingly to the dynamics of FLRW models and the vacuum regions will remain unchanged. The formation of black holes will start from the interior, as the smallest dust region has the biggest density. The result of the entire structure collapsing would be a single Schwarzschild region with mass $M_0$, as seen by the external observer.
\section{Propagation of the radial photons}
\begin{figure}[h]
\centering
\includegraphics[scale=0.35]{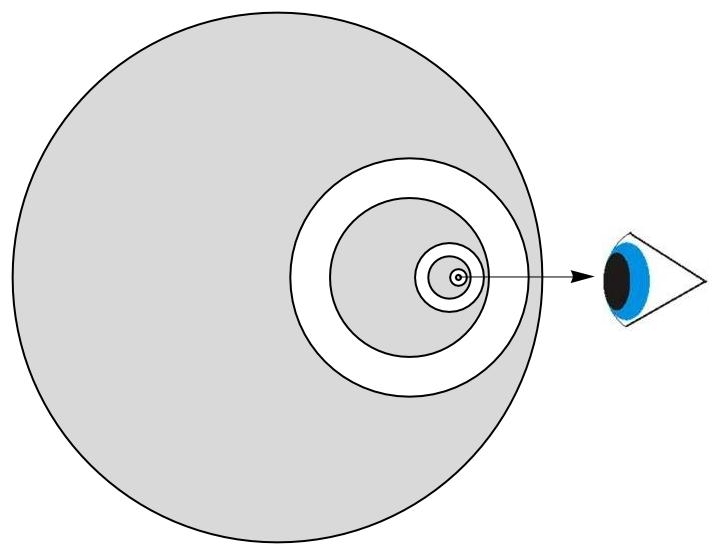}
\caption{Schematic representation of photons trajectory.}  \label{fig: fig2}
\end{figure}

One of the ways for the external observer to asses whenever the self-similar structure can be described using the weak field limit is to study the redshift of photons coming from the center. If it will be small, then this limit should suffice to describe this system. For the simplicity, we consider only radial photons. The geodesic equation for this case written in the LTB coordinates is
\begin{equation} \label{eq: redfor}
\frac{\mathrm{d}r}{\mathrm{d}t} = \frac{\sqrt{1-2E}}{R,_r}.
\end{equation}
The starting point and the direction of the photon is chosen in such a way that it will propagate from the center of the smallest ball of dust to the exterior region along the shortest path. From the axial symmetry of the structure this trajectory, depicted on the Fig. \ref{fig: fig2}, is a geodesics.

\begin{figure}[h]
\centering
\includegraphics[scale=0.35, center]{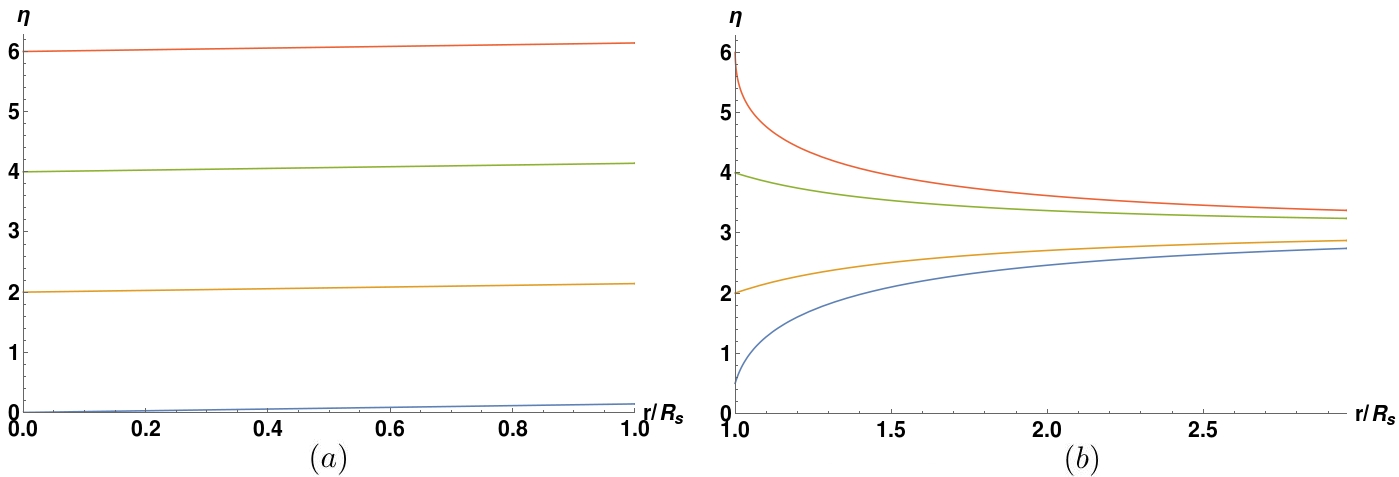}
\caption{Dependence of cycloidal time on radial coordinate for geodesics with varying initial time for the dust (a) and Schwarzshild (b) regions, where $\epsilon = 0.01$.}  \label{fig: fig3}
\end{figure}

In order to compute the redshift of those photons, we use the Bondi formula \cite[Chapter 18]{krasinski},
\begin{equation}
\frac{1}{1+z} \frac{\mathrm{d}z}{\mathrm{d}r} = \frac{R,_{rt}}{\sqrt{1+2E}},
\end{equation}
where $z$ is the desired quantity.
\\
\indent
The description of such propagation is difficult due to the exponential change in parameters describing the structure. As we move closer to the center, the size of dust and vacuum regions are decreasing, whereas densities will increase. In order to overcome this difficulty, we introduce a description using the dimensionless parameters. The main advantage of this procedure is the ability to describe the full propagation through $N$ pairs of FLRW and Schwarzschild regions as a propagation through one pair, repeated $N$ times.
\\
\indent
The first dimensionless parameter, cycloidal time $\eta$, comes from the solution of (\ref{eq: baseq}) for the positive $E$,
\begin{equation}
 \begin{aligned} \label{eq: Reta}
R(r,t)&=-\frac{M}{2E} \big( 1-\cos \eta \big), \\
\eta - \sin \eta &= \frac{\left(-2E \right) ^{\frac{3}{2}}}{M}\left(t-t_{b}\right),
\end{aligned}
\end{equation}
with the Big Bang function $t_b(r)$, which determines the time distance of $r=\mathrm{const}$ surface from the initial singularity. We can define $\eta$ separately for dust and vacuum regions,
\begin{enumerate}[I]
 \item Dust region
 \begin{equation} \label{eq: etako}
 \begin{aligned}
\eta&=\arccos\left(1-\frac{2R}{r}\right), \\
\eta_n &- \sin \eta = \frac{2}{R} \sqrt{\frac{2M}{R}} \left( t-t_{bI} \right),
\end{aligned}
 \end{equation}
  \item Vacuum region
 \begin{equation} \label{eq: etasch}
 \begin{aligned}
\eta&=\arccos\left(1-\frac{2R}{r}\right), \\
\eta &- \sin \eta  = 2\sqrt{\frac{2M}{r^{3/2}}}\left( t-t_{bII} \right),
\end{aligned}
 \end{equation}
 \end{enumerate}
 with the initial value $\eta=\pi$ for $t=0$, which determines the $t_{bI}$ and $t_{bII}$.
 \\
\indent
The second parameter is the angular radius $\Lambda_1$ which defines size of the embedded pair in the bigger dust region.
\\
\indent
The full propagation through $N$ pairs can be thought of in terms of the mapping
\begin{equation} \label{eq: mapping}
F(\eta_i,z_i) = (\eta_f, z_f),
\end{equation}
which takes the cycloidal time and redshift of photon entering the dust region as an input and maps it to the same quantities at the matching surface between the Schwarzschild and the bigger FLRW region. After applying (\ref{eq: mapping}) $N$ times, we will obtain the redshift of the photons at the boundary of the whole structure as a function of the initial quantities defined in the smallest ball of dust.
\\
\indent
In order to describe the mapping $F$ in a proper way, a dependence of cycloidal time on the $r$ coordinate for null geodesics has to be derived. Every such trajectory has the tangent vector $V$ of the form
\begin{equation}
V=\partial_r+\frac{\mathrm{d}t}{\mathrm{d}r} \partial_t.
\end{equation}
We are looking for $V(\eta)$. Combining the (\ref{eq: redfor}), (\ref{eq: etako}) and (\ref{eq: etasch}) we have
\begin{equation} \label{eq: veta}
 \begin{aligned}
 V(\eta)=&-\sqrt{\frac{-2E}{1+2E}}\bigg( \log \bigg(\frac{M}{-2E}\bigg)\bigg)_{,r}+ \\
 +&\frac{\eta-\sin \eta-\pi}{1-\cos\eta}\bigg( \log\bigg(\frac{\big( -2E \big)^{3/2}}{M}\bigg)\bigg)_{,r} \bigg(1-\sqrt{\frac{-2E}{1+2E}}\frac{\sin \eta}{1-\cos \eta} \bigg).
 \end{aligned}
\end{equation}
As it turns out, the solution of (\ref{eq: veta}) is linear in the dust region, where the second part vanishes. This is due to the fact that cycloidal time $\eta$ is the natural generalization of cycloidal time defined for FLRW metric. In the vacuum region, the second part of (\ref{eq: veta}) has an effect on the monotonicity of $\eta$ (Fig. \ref{fig: fig3}).

\begin{figure}[h]
\centering
\includegraphics[scale=0.26, center]{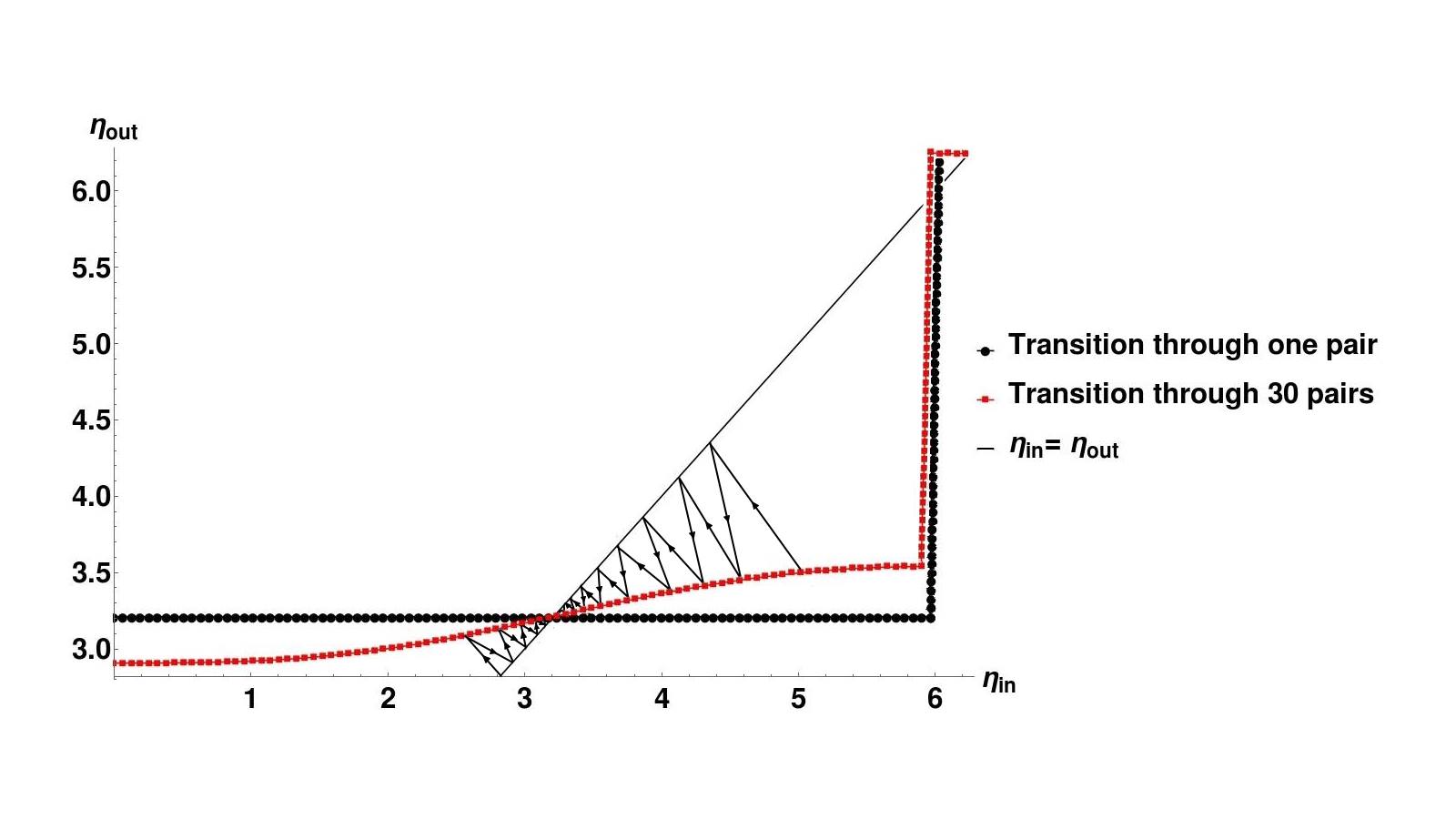}
\caption{Relationship between the initial ($\eta_i$) and final ($\eta_f$) cycloidal time for propagation thorugh one and 30 pairs of regions. Parameters describing the configurationation are $\epsilon=0.01$, $\Lambda=\pi/2$ and $\Lambda_1=\arcsin 0.4$. Attractor is defined by the $\eta_i = \eta_f$.}  \label{fig: fig4}
\end{figure}

\begin{figure}[h]
\centering
\includegraphics[scale=0.3, center]{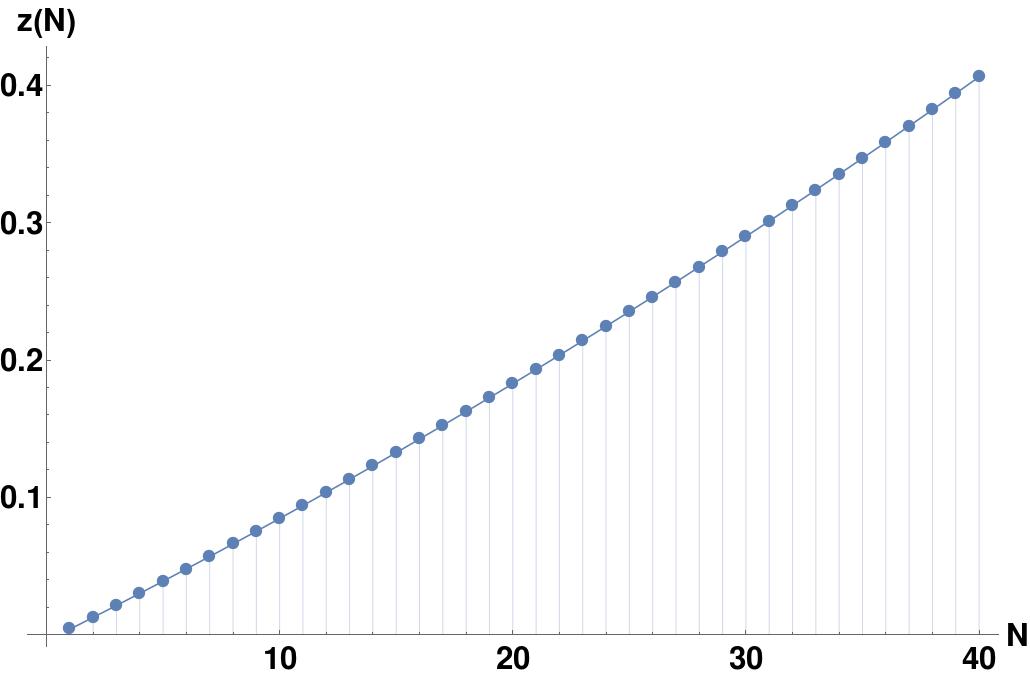}
\caption{Redshift $z$ as a function of number of embedded pairs, $N$.}  \label{fig: fig5}
\end{figure}

We can use solution of (\ref{eq: veta}) to describe the part of the mapping $F$ with the change of cycloidal time (Fig. \ref{fig: fig4}). The discontinuity of $\eta$ is related to the fact that photons which are emitted after the dust region has collapsed and formed black hole will not escape to the external observer. Another feature is the existence of the attractor, both for the propagation through one pair as for the 30 pairs. It means that the observer will measure all of the emitted photons in the short period of time.
\\
The existence of the attractor is also important for the measured redshift and manifests itself in a fact that if a photon will not be caught in the first collapsing ball of dust, then this quantity will be increasing with the number of pairs, $N$. In order to put it in a quantitative way, we should express $z$ as a function of cycloidal time, thus getting the full form of mapping $F$. Using again the (\ref{eq: redfor}), (\ref{eq: etako}) and (\ref{eq: etasch}), we end up with the differential equations,
\begin{equation}
\frac{1}{1+z}\frac{dz}{dr}=\frac{\sin \eta}{1-\cos \eta}\sqrt{\frac{2 M}{R^3-2 M r^2}},
\end{equation}
for the dust regions, and
\begin{equation}
\begin{aligned}
\frac{1}{1+z}\frac{dz}{dr}=&-\sqrt{\frac{2M}{1-\frac{2M}{r}}}\frac{r^{-3/2}}{2-2 \cos\eta}\left(-\sin \eta + \left(\frac{3 \sin \eta }{{(1- cos \eta)^2}}-\frac{3 \cos \eta }{1- \cos \eta }\right)(\eta - \sin \eta  - \pi) \right),
\end{aligned}
\end{equation}
in the Schwarzschild region. Using this relations, we can compute the dependence of redshift on the number of embedded pairs, $N$ (Fig. \ref{fig: fig5}). As it turns out, the dependence $z(N)$ can be described with a linear function,
 \begin{equation}
z(N) = \epsilon x N + O(\epsilon^2), 
 \end{equation}
 where $x \approx 1$. This means that although the one pair of regions can be described by a weak-field approximation of general relativity, the whole structure can manifest strong-field effects if the number $N$ is big enough.

\end{document}